\documentclass[a4paper, prl, twocolumn, showpacs]{revtex4}

\usepackage{amssymb}
\usepackage{amsmath}
\usepackage{epsfig}
\usepackage{color}
\usepackage{graphics, graphicx}
\usepackage{bbold}
\usepackage{psfrag}
\usepackage{mathcomp}

\begin{document}

\author{Michiel Snoek}
\author{H. T. C. Stoof}
\affiliation{Institute for Theoretical Physics, Utrecht University, Leuvenlaan
4, 3584 CE Utrecht, The
Netherlands}

\title{Vortex-lattice melting in a one-dimensional optical lattice}

\begin{abstract}
We investigate quantum fluctuations of a vortex lattice in a
one-dimensional optical lattice. Our method gives full access to all the modes
of the vortex lattice and we discuss in particular the Bloch bands of the
Tkachenko modes.
Because of the small
number of particles in the pancake Bose-Einstein condensates at every site of
the optical lattice, finite-size effects become very
important. Therefore, the 
fluctuations in the vortex positions are inhomogeneous and the melting of the
lattice occurs from the outside inwards. 
Tunneling between neighbouring pancakes substantially reduces the
inhomogeneity as well as the size of the fluctuations.
\end{abstract}

\maketitle
{\em Introduction.} --- Very rapidly rotating ultracold bosonic gases have been
predicted to form 
highly-correlated quantum states \cite{Wilkin98, Cooper01, Ho02}.
In these states, the Bose-Einstein condensate has been completely depleted by
quantum fluctuations, and quantum liquids appear with excitations that can carry
fractional statistics. Some of these states have been identified with (bosonic)
fractional quantum Hall states \cite{Cooper01, Paredes02, Regnault03}. It is by
now a long standing goal to observe the experimental signatures of these very
interesting states in the context of ultracold quantum gases. The conditions for
these states to form have been expressed in the requirement that  the
ratio $\nu=N/N_v$ of the number of atoms $N$ and the number of vortices $N_v$,
should be smaller than some critical value $\nu_c$. The ratio $\nu$ plays the
role of the filling factor and estimates for the
critical $\nu_c$ are typically around $8$ \cite{Cooper01, Sinova02}. These
estimates are made for infinitely large systems.

\begin{figure}[b]
\vspace{-.7cm}
\includegraphics[scale=.25, angle=270, origin=c]{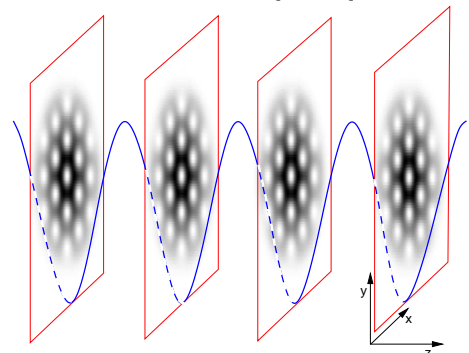}
\vspace{-1.1cm}
\caption{(Color online) Setup in which the melting of the vortex lattice is
studied. 
An optical lattice along the $z$-direction (indicated in blue) divides the
condensate 
into pancake condensate which are coupled by tunneling processes.}
\label{cartoon}
\end{figure}

However, observed filling factors are up till now always greater than 100, where
almost
perfect hexagonal lattices form and no sign of melting can be seen
\cite{Schweikhard04}. These experiments are carried out with Bose-Einstein
condensates consisting of typically $10^5$ particles, whereas the maximum number
of vortices observed is around $300$. Decreasing the number of particles results
in loss of experimental signal, whereas the number of vortices is limited by the
rotation frequency that has to be smaller than the transverse trapping
frequency. Adding a quartic potential, which stabilizes the condensate also for
rotation frequencies higher than the transverse trap frequency, has until now
not improved this situation \cite{Bretin04}, although it has opened up the
possibility of forming a giant vortex in the center of the cloud
\cite{Fetter01}. 
In recent work, however, it was realized that quantum
fluctuations
of the vortices can be greatly enhanced, without loosing experimental signal, by
using a one-dimensional optical
lattice  \cite{Martikainen03}. 
The optical lattice divides the Bose-Einstein condensate into a stack of
two-dimensional pancake condensates that are weakly coupled by tunneling as
schematically 
shown in Fig. \ref{cartoon}. The number of particles in a single
pancake is much smaller than in a Bose-Einstein condensate in a harmonic
trap, and therefore the fluctuations are much 
greater.
Moreover, by varying the coupling between the pancakes, it is
possible to
study the dimensional crossover between decoupled two-dimensional melting, and
the strong-coupling limit where the three-dimensional situation is recovered.
Recently, the density profiles for quantum Hall liquids in such a geometry have
also been calculated \cite{Cooper05}. 
Because of the small number of particles in each pancake shaped Bose-Einstein
condensate, finite-size
effects become very pronounced in this setup. In particular, 
the critical filling factor for the melting of the lattice $\nu_c$ changes
compared to the  
homogeneous situation. Moreover, melting is not expected to occur homogeneously
but starts at the outside and then gradually moves inwards as the rotation speed
increases. Therefore,  phase coexistence is expected, where a vortex crystal is
surrounded by a vortex liquid. 
In this Letter we study this interesting physics by investigating the quantum
fluctuations
of the vortex lattice for realistic numbers of particles and vortices. To decide
whether or not the vortex 
lattice is melted we use the Lindemann criterion, which in this case has
to be applied locally. 

Besides the possibility of observing highly-correlated quantum states after the
vortex lattice has melted, also the vortices themselves have very interesting
properties. For moderately
rotating gases, the 
vortices are observed to order in a Abrikosov lattice \cite{Abo01, Madison00}.
The lowest mode of the lattice is the so-called Tkachenko mode
\cite{Tkachenko66}, which recently has been investigated experimentally
\cite{Schweikhard04, Coddington03} and theoretically \cite{Baym03, Mizushima04,
Baksmaty04}. Since the study of the fluctuations of the vortex lattice requires
full information on all the modes of the lattice, we also can identify this
mode. Because of the optical lattice, this mode actually becomes a Bloch band
and we also determine the dispersion of the Tkachenko modes in the axial
direction of the optical lattice.

{\em Lattice vibrations.} --- 
The action describing the system in the rotating frame is given by $S= \int dt
\int d^3 {\bf x} \; \mathcal{L}({\bf x}, t)$, with the Lagrange density given
by
\begin{equation}
\mathcal{L} =  \Psi^* 
\left(i \hbar \partial_t 
+ \frac{\hbar^2 \nabla^2}{2m} 
- V({\bf x}) 
+ \Omega L_z   - \frac{g}{2} |\Psi|^2 \right)\Psi,
\end{equation}
where $\Psi ({\bf x}, t)$ is the Bose-Einstein condensate wavefunction, $m$
is the mass of the
atoms, 
$\Omega$ is the rotation frequency, $L_z= i \hbar (y \partial_x - x
\partial y)$ is the angular momentum operator,
and $g= 4 \pi \hbar^2 a/m$ is
the interaction strength, 
with $a$  the three-dimensional scattering length. The combination of an optical
lattice in the axial direction and harmonic confinement in the radial direction
gives an external
potential 
$V({\bf x}) = V_z \cos^2 (\lambda z/2) + \tfrac{1}{2} m \omega_\perp^2
(x^2+y^2)$, where
$\lambda$ is the wavelength of the 
laser producing the optical lattice.
Assuming that the optical lattice is deep enough we can perform a tight-binding
approximation and
write 
$\Psi({\bf x},t)= \sum_i \Psi (z-z_i) \Psi_i(x,y, t)$, where $\Psi(z)$ is
chosen
to be the lowest Wannier function
of the optical lattice. In this way we obtain the Lagrange density
\begin{eqnarray}
\mathcal{L} \!\! &=& \!\! \sum_i \Psi_i^* (i \hbar \partial_t +
\frac{\hbar^2 \nabla^2}{2m} 
- \frac{ m \omega_\perp^2}{2}  r^2
+ \Omega L_z - \frac{g'}{2} |\Psi_i|^2
)\Psi_i
\nonumber \\&&
+ t \sum_{\langle i j \rangle} \Psi_i^* \Psi_j, 
\end{eqnarray} 
where $\langle i j\rangle$ denotes that the sum is taken over nearest
neighbouring sites.
We have defined the interaction coefficient $g'=4 \pi \hbar^2 a/\sqrt{2
\pi}\ell_z m$ and hopping amplitude
$t=4 V_z^{3/4} E_z^{1/4} \exp[-2 \sqrt{V_z/E_z}]$,  where $E_z=2 \pi^2 \hbar^2
/\lambda^2 m$ is the recoil energy associated with the optical lattice and  $l_z
= \lambda/2\pi (V_z/E_z)^{1/4}$ is the harmonic length associated with the
optical lattice. 


Melting is only expected for a Bose-Einstein condensate that is weakly 
interacting. The wavefunction can then be taken to be part of the lowest 
Landau level. That is to say we consider wavefunctions of 
the form $\prod_i (z-z_i)\exp[-|z|^2/2]$, where $z = (x+ i y)/\ell$, 
$z_i = (x_i+ i y_i)/\ell$ and $(x_i, y_i)$ is the position of the 
$i^{\rm th}$ vortex. Here $\ell$ is the 'magnetic length'.  To 
increase the validity of this study, we use $\ell$ as a variational parameter
instead of fixed it to the radial harmonic length, such that our results are
also valid for stronger interactions \cite{Mueller04}. 
The associated frequency is $\omega=\hbar/m \ell^2$. 
From now on distances are rescaled by $\ell$,
frequencies are scaled by the radial trappig frequency $\omega_\perp$, 
and we define a dimensionless interaction strength by means of 
$U= N g'/4 \pi \ell^2 \hbar \omega_\perp=  Na/ \sqrt{ 2
\pi}\ell_z$. In the calculations we take the scattering length of $^{87}$Rb,
$\lambda = 700$ nm and $V_z/E_z=16$, which gives $U=25N$.
The on-site energy part of the Lagrange density depends in this approximation
only on the particle density and becomes
$(\omega + 1/\omega - 2 \Omega)
 r^2 n({\bf r})/2
+ 2 \pi \omega U  n^2( {\bf r}).
$
The lowest Landau level wavefunctions and, therefore, also the atomic density,
are fully determined by the location of the vortices. To consider the quantum
mechanics of the vortex lattice we, therefore, replace
the functional integral over the condensate wavefunctions by a path integral
over the vortex positions.
\begin{figure}
\includegraphics[scale=.8]{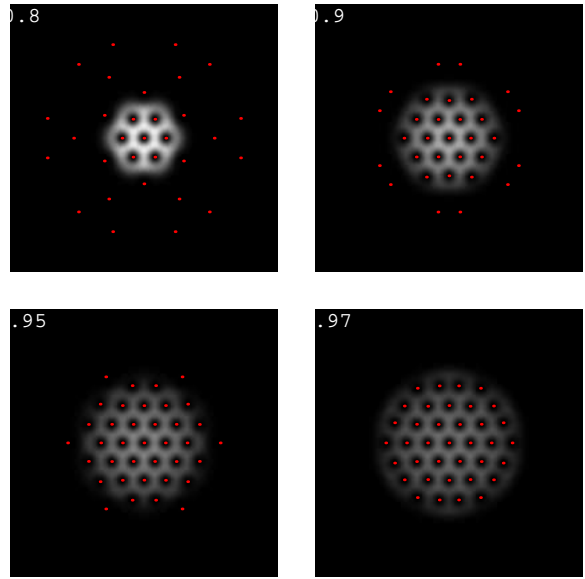}
\caption{(Color online) Classical vortex lattice and density profile 
for rotation frequencies $\Omega/\omega_\perp=.8$, $.9$, $.95$ and $.97$. Here
$U=10$, which corresponds to $N=250$.
White means high density, black low density. The vortex positions are indicated
by a red dot, such that
also the vortices outside the condensate are visible.}
\label{lattice}
\end{figure}

Next we want to determine the quadratic fluctuations around the Abricosov
lattice. To do so, we first have to find the classical groundstate. We
calculated this
groundstate for up to 37 vortices. For small numbers of vortices, the
groundstate is
distorted from the hexagonal lattice \cite{Aftalion05}. In
general, there are also
vortices far outside the condensate. The coarse-grained atomic density is
well approximated by a Thomas-Fermi profile \cite{Cooper04}. For fixed interaction $U$ and
different rotation frequencies pictures
of the classical groundstate are given in Fig \ref{lattice}.
Then we study the quadratic fluctuations by expanding the action up to second
order in the fluctuations \cite{Kim04}.
This yields an action of the form
\begin{equation}
S=\sum_i {\bf u}_i \cdot ({\bf T} i \partial_t - {\bf E}) \cdot {\bf u}_i - t
\sum_{\langle i
j \rangle} {\bf u}_i \cdot {\bf J} \cdot {\bf u}_j,
\end{equation} 
where ${\bf u}_i=(\delta x_{1, i},\delta y_{2, i},\delta x_{2, i}, \delta y_{2,
i}, \ldots) \equiv(\dots, {\bf u}_{ni}, \ldots)$ is the total displacement
vector of all the point vortices on site $i$, and $\bf T$, $\bf E$ and $\bf J$
are matrices depending on $\Omega$, $U$, and the classical lattice positions. 
To diagonalize this action along the $z$-axis, we perform a Fourier
transformation to obtain
\begin{equation}
S=\sum_k {\bf u}_k^*  \cdot ({\bf T} i \partial_t - {\bf E} - t(1-\cos [k
\lambda/2] {\bf J} ) \cdot {\bf u}_k  
\end{equation} 
Finally, we completely diagonalize this action by the transformation ${\bf
v}_k={\bf P}_k {\bf u}_k$ such that the action becomes $
S=\sum_{k, \alpha} v_{k\alpha}^*  (i \partial_t - \omega_\alpha(k) )  v_{k
\alpha}  
$
where $\omega_\alpha(k)$ are the mode frequencies of the vortex lattice. This
means that the $v_{k\alpha}$, where $k$ labels the momentum in the $z$-direction
and $\alpha$ labels the mode, correspond to bosonic
operators 
with commutation relation $[v_{k\alpha}, {v_{k'\alpha'}}^\dagger] = \delta_{kk'}
\delta_{\alpha \alpha'}$. This allows us to calculate the expectation value for
the
fluctuations in the vortex positions, but also for 
the correlations between the various point vortices. 

\begin{figure}
\psfrag{omega(k)}{$\omega(k)/\omega_\perp$}
\psfrag{k}{$k \lambda/2$}
\includegraphics[scale=.66]{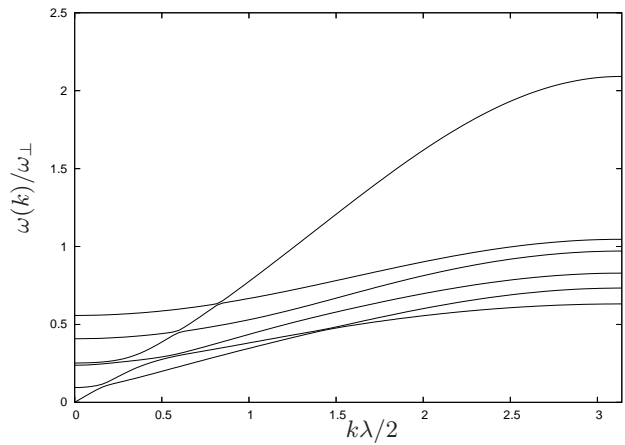}
\caption{Dispersion of the Tkachenko modes along the direction of the optical
lattice. For this plot the parameters where chosen as: $U=10$, $t=1/10$, and $\Omega
= .97$. One gapless linear mode and five
gapped tight-binding like modes can be identified. The gapless mode is the acoustic Goldstone
mode associated with the broken $O(2)$-symmetry due to the presence of the
vortex lattice.}
\label{dispersion}
\end{figure}

{\em Tkachenko modes.} ---
The Tkachenko modes are purely transverse modes of the vortex lattice.
In a harmonic trap with spherical symmetry they 
become modes which are purely angular.  In the radial direction their spectrum
is discretized, because of the finite lattice size. The number of radial
Tkachenko modes equals the number of vortex rings. For 37 vortices 6 Tkachenko
modes can be identified. A close comparison with continuum theory for a
finite-size system, where also a discrete spectrum was found \cite{Sonin05}, is
possible but beyond the scope of this Letter.
Moreover, the Tkachenko modes also have a dispersion in the $z$-direction.
Without the optical lattice some aspects of these modes were recently investigated
\cite{Chevy05}.
For typical parameters this dispersion is plotted in Fig. \ref{dispersion}. 
As is clearly visible, there is one gapless mode, which is linear at long
wavelengths, while the other modes 
are roughly just tight-binding like. Moreover, various avoided crossings
between these modes are clearly visible. The gapless mode is the Goldstone mode
associated with the spontaneously broken rotational $O(2)$-symmetry due to the
presence of the vortex lattice. 
When the tunneling rate is very small, the gapped modes have exactly a
tight-binding dispersion and the gapless mode gets a dispersion proportional to
$\sin[k \lambda /2]$. This can be understood by observing that in this case the
modes are decoupled and the Hamiltonian for the gapless mode reduces to the
Josephson Hamiltonian $\mathcal{H}= - E_c \sum_i \partial^2/\partial \phi_i^2 +
E_J \sum \cos(\phi_i - \phi_j)^2$, which indeed has this dispersion upon
quadratic expansion.

It is interesting to note that for a small rotation frequency, which implies a small vortex lattice, the Tkachenko
modes are not the lowest-lying modes. For $U=10$, a Tkachenko
mode becomes the lowest-lying gapped mode when $\Omega>.978$, but there are many
modes in between the second and the third Tkachenko mode.
This confirms the expectation that increasing the vortex lattice will bring down
the Tkachenko spectrum more and more.

\begin{figure}
\psfrag{N_v}{$N_v$}
\psfrag{Omega}{$\Omega/\omega_\perp$}
\psfrag{R_cr/R}{ $R_{\rm cr} / R$}
\includegraphics[scale=.66]{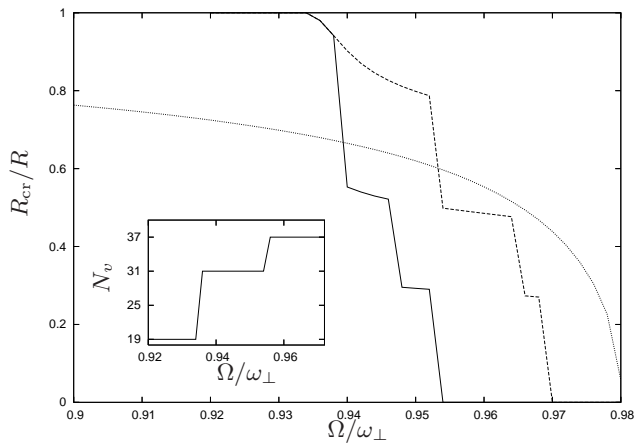}
\caption{Crystal radius $R_{\rm cr}$ normalized to the condensate radius $R$ as
a function
of the rotation frequency for $N= 300$, $U=10$. The full and dashed line are for
$t=0$ and $1/100$, respectively.
The dotted line is the result of a local-density
approximation. In the inset the number of vortices $N_v$ that is within the
condensate is
plotted as a function of the rotation frequency.}
\label{melting}
\end{figure}

{\em Vortex-lattice melting.} ---
Quantum fluctuations of the vortices ultimately result in
melting of the vortex lattice.
To decide wether or not the lattice is melted, we use the Lindemann criterion,
which in this inhomogeneous situation has to be applied
locally.
The Lindemann criterion means that the lattice is melted, when the ratio 
$\langle u_{ni}^2 \rangle/\Delta_{ni}^2$ of the displacement fluctuations
of vortex $n$ at site $i$ and the square of the average distance to the
neigbouring vortices $\Delta_{ni}$ exceeds
a critical value $c_L^2$, which is known as the Lindemann parameter.  
For easy comparison with previous work on this topic, we use the value $c_L^2 =
.02$. 
Because the coarse-grained particle density decreases with the distance to the
origin, 
vortices on the outside are already melted, while the inner part of the crystal
remains solid. 
Therefore a crystal phase in the inside  coexists with a liquid
phase on the outside. In Fig. \ref{melting} we compute the radius of the crystal
phase $R_{\rm cr}$ normalized 
to the condensate radius $R$, as a function of the rotation frequency for fixed
numbers of particles and an interaction strength $U$, and for various hopping
strengths $t$. 
Here we define  the condensate radius $R$ as the radius for with the angularly
averaged density drops below $.003$. The crystal radius $R_{\rm cr}$ is defined
as the radius of the innermost vortex ring that is melted according to the
Lindemann criterion. When according to this definition  $R_{\rm cr}>R$, we set
the crystal radius equal to the condensate radius, i.e., $R_{\rm cr}=R$.  
The ratio $R_{\rm cr}/R$ shows discrete steps because of the ring-like structure
in which the vortices order themselves. 
We compare this with a simple local density calculation, where the criterion
$N/N_v=n(r)/n_v( r)=8$ derived in Ref. \cite{Sinova02} is applied
locally, using a Thomas-Fermi density profile to describe the coarse-grained
atomic density. From the on-site energy we extract that the variational
parameter $\ell$ is set by the condition $ \omega=  \Omega$, and the 
Thomas-Fermi radius is given by $R^4=16  \Omega^2 U /(1 -  \Omega^2)$. When
taking a constant vortex density $n_v ( r) = 1/\pi \ell^2$, the crystal
radius normalized to the Thomas Fermi radius becomes
$R_{\rm cr}/R_{\rm TF} = \sqrt{1 - 4 R^2/N} = \sqrt{1- 16/N \sqrt{
\Omega^2 U/(1- \Omega^2)}}$. For comparison this line is plotted in Fig.
\ref{melting}. As expected for a
finite system, the melting occurs considerably earlier than predicted by the
local-density theory.

When the tunneling between pancakes is turned on, the fluctuations will also be
coupled in the  axial direction. This decreases the fluctuations in the vortex
displacements because the stiffness of the vortices increases, and melting
occurs for higher rotation frequencies,
as is visible in Fig. \ref{melting}.
To determine the presence of crystalline order in the axial direction, we
calculate
$\langle e^{ i {\bf q} \cdot ({\bf u}_{Ni} - {\bf u}_{nj})} \rangle$, which is
related to the structure
factor. For the central vortex we always find long-range order, while for the
other vortices, we obtain an algebraic decay. This is in agreement with the
expectation for a one-dimensional system at zero temperature. 
Note that  the quantitative results  in Fig. \ref{melting} depend
on the value of the Lindemann
parameter. Changing this
value shifts the curves, but the qualitative features
remain the same. 


The liquid surrounding the crystal has rotational symmetry.
Experimentally this can be observed, by noting that
in the liquid the vortices are no longer individually visible, while in the crystal they are
still visible, although their positions are broadened by the fluctuations.
Theoretically it is a challenging problem
to describe the coexisting crystal-liquid. This will allow to decide on the
occurrence of melting based
on energy considerations and thus shed more light on
the accuracy of the application of the Lindemann criterion in this
inhomogeneous situation. 
In the future we want
to make a closer connection with continuum theory. 
Finally we want to extend the analysis to include a quartic potential, in which
case giant vortex formation is prediction. 
In this case the liquid can also form in the center, instead
of at the outer edge.
Finally we also want to investigated the melting at nonzero temperatures, which
is
experimentally relevant, because the zero-temperature limit is difficult to reach
\cite{Cornell}.

We thank Masud Haque, Jani Martikainen,  and Nigel Cooper for
helpful discussions. 
This work is supported by the Stichting voor
Fundamenteel Onderzoek der Materie (FOM) and the Nederlandse
Organisatie voor Wetenschappelijk Onderzoek (NWO).

\end{document}